\documentclass[prl,superscriptaddress,twocolumn,preprintnumbers,a4,showpacs,floatfix]{revtex4}
\usepackage{amsmath, amssymb, psfrag, tabls, dcolumn, bm, color}
\usepackage[sort&compress]{natbib}
\usepackage[pdftex]{graphicx}
\usepackage{graphicx}
\usepackage{textcomp}

\newcommand\smallurl[1]{{\tiny \url{#1}}}

\newcommand{\be}{\begin{equation}}
\newcommand{\ee}{\end{equation}}
\newcommand{\bea}{\begin{equation*}}
\newcommand{\eea}{\end{equation*}}
\newcommand{\ba}{\begin{array}}
\newcommand{\ea}{\end{array}}
\newcommand{\beqa}{\begin{eqnarray}}
\newcommand{\eeqa}{\end{eqnarray}}
\newcommand{\beqaa}{\begin{eqnarray*}}
\newcommand{\eeqaa}{\end{eqnarray*}}

\newcommand{\matr}{\left( \begin{array}}
\newcommand{\ematr}{\end{array} \right)}

\newcommand{\kb}{\mbox{\boldmath $k$}}

\newcommand{\lsim}{{\;\raise0.3ex\hbox{$<$\kern-0.75em\raise-1.1ex\hbox{$\sim$}}
\;}}
\newcommand{\gsim}{{\;\raise0.3ex\hbox{$>$\kern-0.75em\raise-1.1ex\hbox{$\sim$}}
\;}}

\usepackage{mdwlist}

\def\bcols{\begin{columns}}
\def\ecols{\end{columns}}
\def\bcol{\begin{column}}
\def\ecol{\end{column}}
\def\bit{\begin{itemize}}
\def\eit{\end{itemize}}
\def\ben#1{\begin{enumerate}[#1]}
\def\een{\end{enumerate}}
\def\colb#1{\begin{columns}\begin{column}{#1}}

\def\cole{\end{column}\end{columns}}

\usepackage{bm}

\begin{document}

\title{Electronic and optical trends in carbon nanotubes under pure bending}

\author{Pekka Koskinen}
\email[email:]{pekka.koskinen@iki.fi}
\address{NanoScience Center, Department of Physics, University of Jyv\"askyl\"a, 40014 Jyv\"askyl\"a, Finland}




%


\begin{abstract}
The high aspect ratio of carbon nanotubes makes them prone to bending. To know how bending affects the tubes is therefore crucial for tube identification and for electrical component design. Very few studies, however, have investigated tubes under small bending well below the buckling limit, because of technical problems due to broken translational symmetry. In this Letter a cost-effective and exact modeling of singe-walled nanotubes under such small bending is enabled by revised periodic boundary conditions, combined with density-functional tight-binding. The resulting, bending-induced changes in electronic and optical properties fall in clear chirality-dependent trend families. While the correct trends require full structural relaxation, they can be understood by one general argument. To know these trends fills a fundamental gap in our understanding of the properties of carbon nanotubes.
\end{abstract}

\maketitle

Carbon nanotubes (CNTs), with an aspect ratio akin to tens of meters long human hair, bend frequently.\cite{ijima_nature_91,falvo_nature_97} In most transmission electron micrographs CNTs appear as curly web of hair, unless placed on a support with special care.\cite{ko_APL_04, pan_JAP_02,song_APL_08,wang_PRB_10} Bending is hard to avoid in experiments completely.

With theory the state of affairs is just the opposite. On one hand, it's easy to model infinitely long and perfectly straight tubes, using Bloch's theorem and periodic boundary conditions. On the other hand, it's usually difficult to model large-scale distortions once the translational symmetry gets broken. Bending has been therefore commonly modeled by finite tubes and thousands of atoms, accessible only by classical potentials.\cite{kutana_PRL_06,yakobson_PRL_96}

However, since CNTs serve mostly as electronic components, bending and other distortion studies ought to incorporate the electronic structure.\cite{pastewka_PRB_09} Sure enough, density-functional theory has been used to study CNT bending, but reliable electronic trends are hard to extract, because tubes are short and bending violent---mostly near or above the buckling limit.\cite{rochefort_PRB_99, tang_JAP_10} Results also depend on how the tubes are bent, whether bending is achieved by forces from external bodies, or by geometrical constraints.

In this Letter a cost-effective way to model tubes under pure bending is enabled by revised periodic boundary conditions (RPBC). In this approach the bending emerges naturally, being caused by boundary conditions alone, and is devoid of spurious effects. While similar approach for carbon nanotubes was used earlier by Dumitric\u a \emph{et al.},\cite{nikiforov_APL_10,dumitrica_JMPS_07} as well as Malola \emph{et al.},\cite{malola_PRB_08b}, they did not study electronic properties. The focus in this work is, therefore, to answer the question: ``How do band gaps and optical properties change as single-walled CNTs get slightly bent?'' The answers to this question, as it will turn out, have clear trends that can be understood with one fundamental argument.

\begin{figure}[t]
\includegraphics[width=6cm]{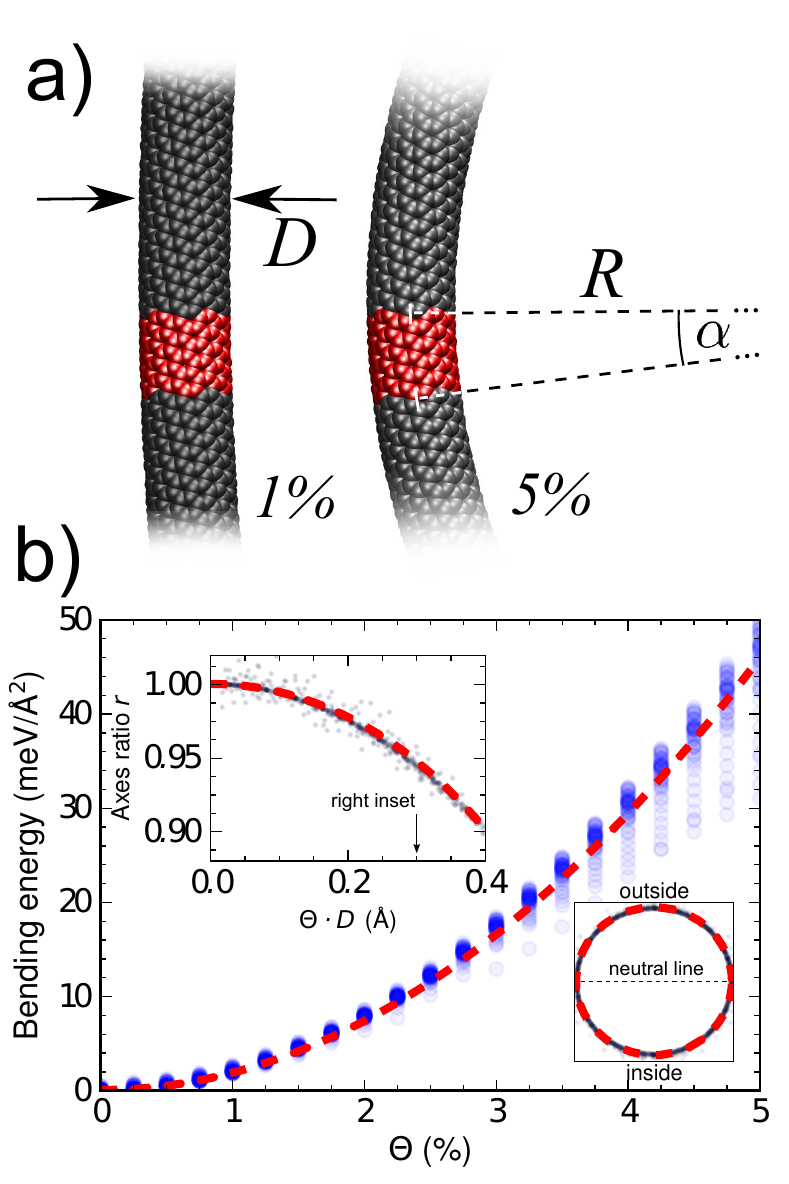}
\caption{(color online) (a) $(10,5)$ CNT, with $D=10.3$~\AA, bent to $\Theta=1$~\% and $\Theta=5$~\%. Red (light gray) atoms denotes the simulation cell, and the symmetry operation is a rotation of an angle $\alpha$ around given origin. $\Theta$ is a universal measure of bending: the impression for the amount of bending for given $\Theta$ is independent of $D$. (b) Elastic bending energy (per unit length and divided by tube diameter) as a universal function; the dashed line is the analytical expression, Eq.(\ref{eq:E}). Right inset: cross-section view of atoms' positions from all tubes, plotted with $D\Theta= 0.3$~\AA. The dashed line is an ellipse with $r=0.949$, as given by Eq.(\ref{eq:r}) with $D\Theta= 0.3$~\AA. The flattening is only somewhat visible when comparing to the perfect square. Left inset: ratio of the two major axes of the flattened cross-section. Dashed line is the analytical expression (\ref{eq:r}).}
\label{fig:first}
\end{figure}

To modeling of bending is done by replacing the usual translation symmetry by a rotation symmetry around a given origin, as clarified in Fig.~\ref{fig:first}a; for details of this RPBC approach, see Ref.~\onlinecite{koskinen_PRL_10}. Since the rotation is a symmetry operation, the modeled system as a whole is, in fact, a huge nanotorus. The amount of bending is measured by the parameter
\begin{equation}
\Theta = \frac{D}{2R},
\end{equation}
where $D$ is diameter and $R$ torus's radius. This parameter is universal, and enables comparing tubes with different diameter. (The sidewall strain on the compressive side is $\varepsilon\approx -\Theta$ and on the tensile side $\varepsilon \approx \Theta$, independent of $D$) I concentrate here on bending around the experimentally relevant $\Theta\sim 0.01$ or $1$~\%\ region, with $5$~\% at maximum;\cite{chang_PRL_07} $1$~\% corresponds to bending a human hair into a loop with diameter of $\sim 1$~cm.

The electronic structure itself is modeled by density-functional tight-binding (DFTB) method,\cite{porezag_PRB_95} using the \texttt{hotbit} code with an RPBC implementation.\cite{koskinen_CMS_09,hotbit_wiki} This method uses first principles theory first to parametrize matrix elements, then to fit pair potentials that give total energies.\cite{koskinen_CMS_09} DFTB describes carbon materials reasonably well, regarding both mechanical and electronic properties.\cite{elstner_PRB_98,popov_PRB_04,popov_NJP_04} Considering these literature records, DFTB should describe the physics of bend CNTs well. Note that the approach to treat pure bending is exact, the sole approximation in this work being DFTB itself.\cite{approx2}

The simulation in practice ran as follows. I selected CNTs with different chiralities $(n,m)$ such that the number of atoms in the unit cell was below $210$, which made some $60$ tubes in total. All the tubes were then bent from $0$~\% to $5$~\% in steps of $0.25$~\% in the following way: First, I fixed $\alpha$ according to the $\Theta$ of interest, using reasonable estimates for the initial geometry (including estimate for $R$). Second, I optimized the tubes to the maximum force of $10^{-5}$~eV/\AA.\cite{bitzek_PRL_06,force-note} Note that the only parameter fixed was $\alpha$; the tube can freely move in radial direction, to find the optimum geometry and radius---the bending is pure. The number of equally spaced $\kappa$-points ($\kb$-point equivalent for rotational symmetry) were $50$~\AA$/L$ for optimizations and $500$~\AA$/L$ for electronic structure analysis, where $L$ is the unit length of the straight tube. While the simulation cells contained $\sim 100$ atoms, the largest corresponding torii would have contained $\sim 10^6$ atoms, giving an obvious motivation for this approach.

Prior to presenting the results for bent CNTs' electronic structure, it's illustrating first to review bent tubes' structural trends.\cite{huhtala_CPC_02} Namely, the inset in Fig.\ref{fig:first}b shows that bent tubes get slightly flat. Flattening is due to two competing effects: flatter tubes have on one hand less energy due to strain, but on the other hand more energy due to increased sidewall curvature. Using thin sheet elasticity theory to calculate the optimum between these competing effects, one gets
\begin{equation}
r = 1-\frac{3}{4\left[ (\lambda/\Theta D)^2-1 \right]}
\label{eq:r}
\end{equation}
for the ratio of the minor and major axes of the resulting elliptical cross-section, where $\lambda=\sqrt{18 \kappa/Y} \approx 1.16\text{~\AA}$, $\kappa=1.6$~eV is graphene's bending modulus and $Y\approx 340$~Pa\AA\ its Young's modulus in two dimensions. Equation~(\ref{eq:r}) shows that thicker tubes flatten more easily, and that tubes with equal $D\Theta$ are equally flattened; the inset of Fig.~\ref{fig:first}b shows all tubes' cross-section with $D\Theta=0.3$~\AA, or $r=0.949$. Since buckling means complete flattening of the tube, there ought to be some critical limit $r_\text{c}$ below which tubes buckle. Earlier calculations have shown that $r_\text{c}\sim 0.65$, which means $D\Theta_\text{c} \sim 0.75$~\AA;\cite{yakobson_PRL_96,huhtala_CPC_02} the tubes here have $\Theta_\text{c}=5\ldots 20$~\%, and hence always $\Theta<\Theta_\text{c}$.

The analytical expression for the elastic energy of tubes flattened this optimum way is
\begin{equation}
E = \frac{\pi}{4} D Y \Theta^2
\label{eq:E}
\end{equation}
per unit length.\cite{arias_PRL_08} The quantity $E/D$, which hence is universal for all tubes, is plotted in Fig.~\ref{fig:first}b. Small-diameter tubes deviate most from this behavior because thin-shell theory is less valid. This structural behavior of CNTs under bending agrees with previous classical simulations (using thousands of atoms),\cite{kutana_PRL_06,tang_JAP_10,huhtala_CPC_02,yakobson_PRL_96} and thus, for its part, validates the modeling approach.

\begin{figure}[b]
\includegraphics[width=6cm]{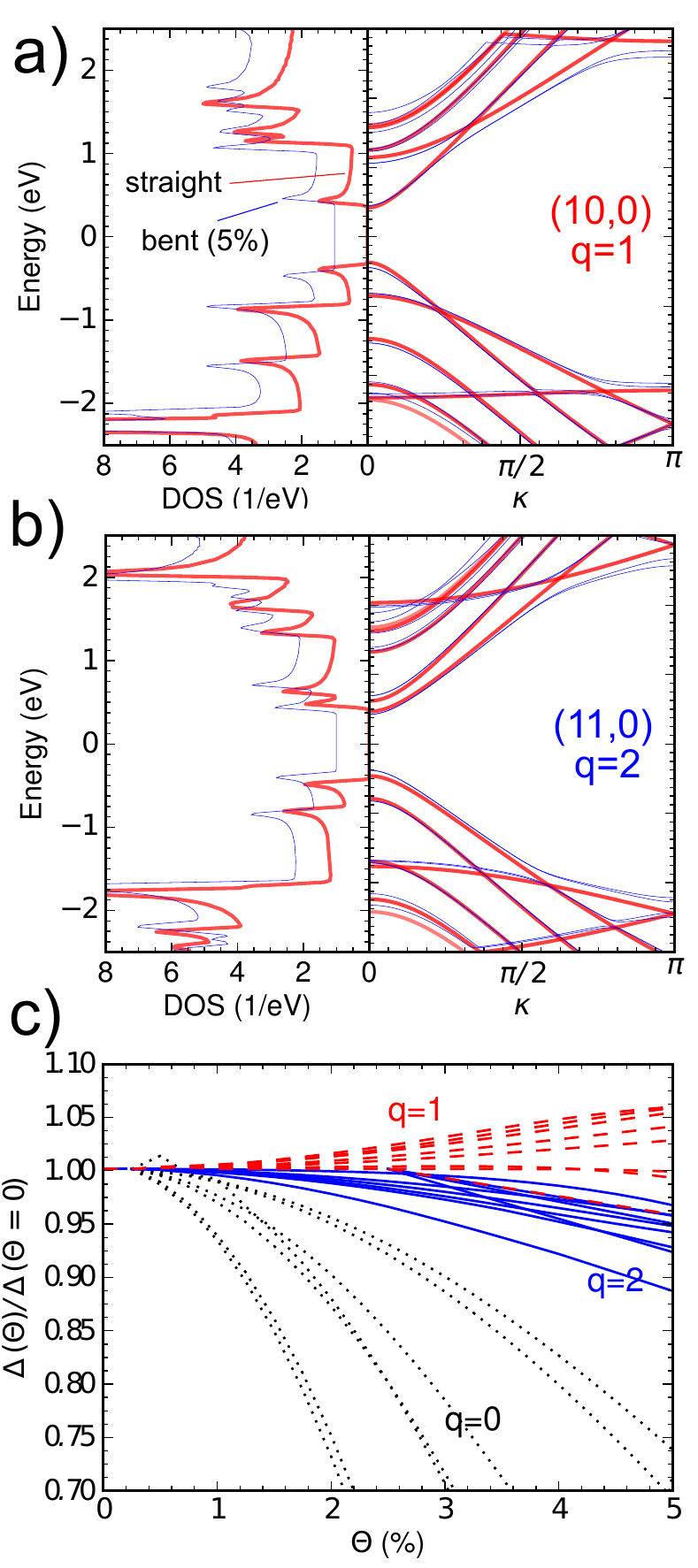}
\caption{(color online) (a) and (b) Density-of states (left) and band-structure (right) for straight and bent $(10,0)$ and $(11,0)$ tubes. (c) The behavior of band gap $\Delta(\Theta)$ as a function of bending depends on tube's $q$-family [defined in Eq.(\ref{eq:q})]. Gaps are renormalized by the gaps of the straight tubes, $\Delta(\Theta=0)$.}
\label{fig:gap}
\end{figure}

Now, let's finally turn the attention to the results in electronic structure. Figure~\ref{fig:gap}a shows the density of states and band structures for straight and bent $(10,0)$ tubes. The first observation is that the bending-induced changes in the electronic structure are gradual. Band extrema---the van Hove singularities---move gradually up and down, while some band anticrossings become more visible. The band gap increases little, yet distinctly. Figure~\ref{fig:gap}b shows the same results for $(11,0)$ tube. The changes due to bending appear similar, but note that the trends are just the opposite: the band extrema move in the opposite direction, and band gap decreases.

The trends in the band gaps $\Delta(\Theta)$ upon bending, summarized in Fig.~\ref{fig:gap}c, is the first main result of this work. The trends fall in families depending on CNT chirality $(n,m)$ as
\begin{equation}
q = (n-m)\text{ mod } 3;
\label{eq:q}
\end{equation}
hence gap decreases for $q=0$ and $q=2$, whereas it increases for $q=1$. For $q\neq 0$ semiconducting tubes having sizeable gaps the relative change in $\Delta$ is smaller than for $q=0$ -tubes with a minigap ($q=0$ tubes are often referred to as being ``metallic''). The $q=0$ tubes are therefore affected the most by bending---the bending can almost make those minigaps vanish.

Trends above have two exceptions, armchair and small diameter tubes, that are excluded from Fig.~\ref{fig:gap}c. First, the electronic structure of armchair tubes ($n=m$), it turned out, are very robust against bending. They remained metallic, with band structure nearly intact. This robustness, explained by symmetry arguments, has been reported earlier.\cite{yang_PRL_00,kane_PRL_97} Second, the electronic structure of tubes with small diameter ($\lesssim 5$~\AA) can change drastically. Those changes are interesting, but they are unsystematic and impossible to understand jointly, and therefore outside the scope of this paper.

\begin{figure}
\includegraphics[width=6cm]{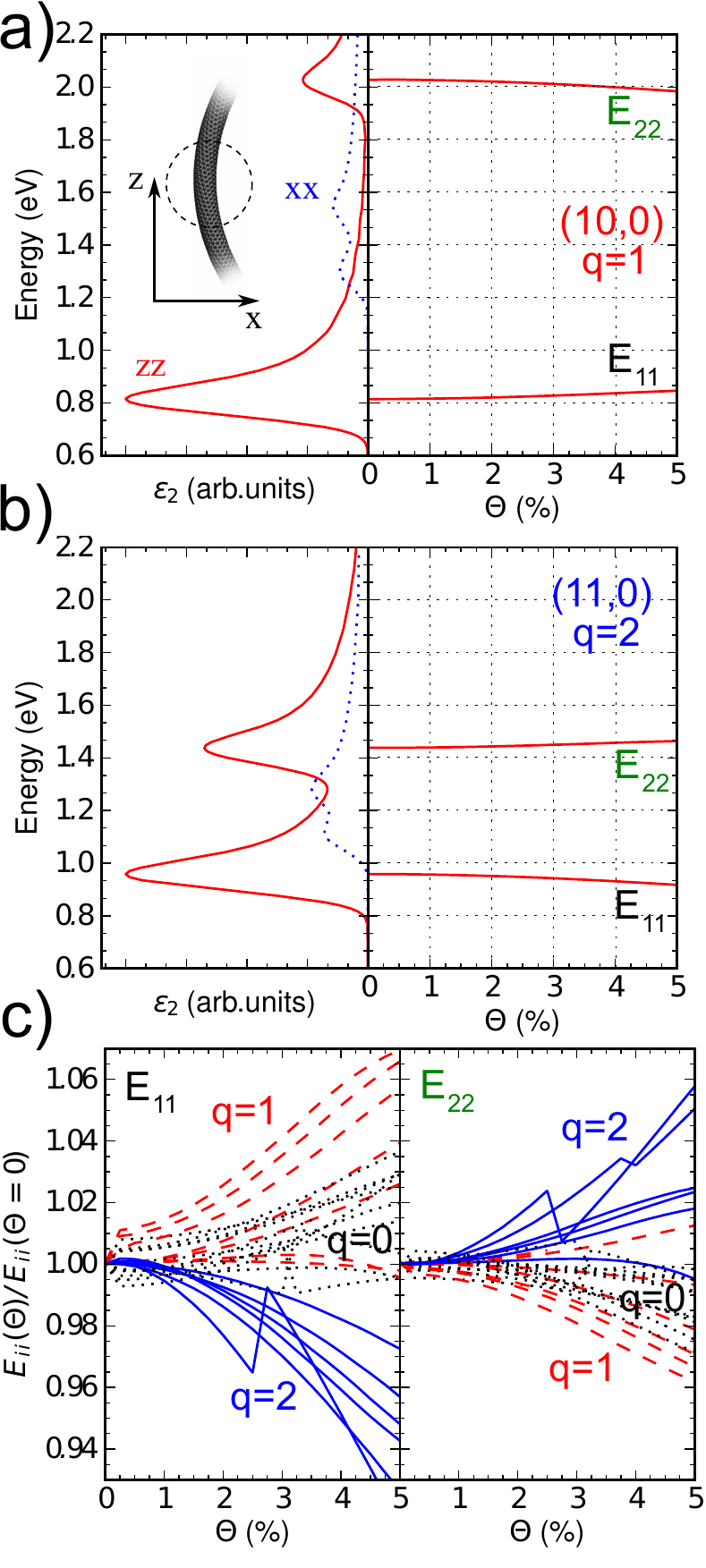}
\caption{(color online) (a) Left: imaginary part of the dielectric function for a straight $(10,0)$ tube; incident and reflected light polarizations are either parallel ($zz$) or perpendicular ($xx$) to tube axis. Right: the effect of bending on the optical $E_{11}$ and $E_{22}$ peaks. (b) The same for $(11,0)$ tube. (c) The behavior of $E_{11}$ (left) and $E_{22}$ (right) transitions depends on the tube $q$-family, Eq.(\ref{eq:q}). Transition energies are renormalized by straight tube energies. The irregularities with $q=2$ are not real, but related to numerical problems in structure optimization; they indicate, however, the sensitivity to properly optimized geometries.}
\label{fig:optical}
\end{figure}

The next trends I shall discuss are optical trends. They are investigated via the imaginary part part of the dielectric function, $\varepsilon_2(\omega)$, which is directly proportional to the optical absorption. The function $\varepsilon_2(\omega)$ is calculated within the random phase approximation.\cite{martin_book} This calculation includes, however, one issue that I wish to discuss first. Assuming both incident and emitted light polarization in the Cartesian direction $\beta$, the calculation will involve matrix elements $\langle \psi_{\kappa,a} |\hat{p}_\beta | \psi_{\kappa,b} \rangle$, where $\hat{p}$ is the momentum operator and where the states $\psi_{\kappa,i}$ are delocalized over the whole torus.\cite{martin_book} While the optical response could indeed be calculated for the nanotorus, the emphasis here is on the optical properties arising from bent fragments of CNTs, not on nanotori as whole, complete objects. I therefore evaluate the matrix element only within the chemical interaction range from the unit cell; since the tube axis reorients very little between neighboring unit cells, fixing the orientation of coordinates makes sense. The resulting $\varepsilon_2(\omega)$ is hence a \emph{local} property: the situation resembles probing and measuring the tube using a laser with a spot size much smaller than the radius of curvature, as sketched in Fig.~\ref{fig:optical}a.


The function $\varepsilon_2(\omega)$ calculated this way for a straight $(10,0)$ tube is then shown in Fig.~\ref{fig:optical}a, on the left. On the right, figure shows further how the first two transition peaks $E_{11}$ (between the first pair of occupied and unoccupied van~Hove singularities) and $E_{22}$ (between the second pair of occupied and unoccupied van~Hove singularities) change as a function of $\Theta$: $E_{11}$ increases and $E_{22}$ decreases in energy, as can be deduced also from the densities of states in Fig.~\ref{fig:gap}a. Figure~\ref{fig:optical}b shows the same for $(11,0)$ tube, and the behavior is, again, the opposite: $E_{11}$ decreases and $E_{22}$ increases in energy.

The trends in transitions $E_{11}$ and $E_{22}$ upon bending, summarized in Fig.~\ref{fig:optical}c, is the second main result of this Letter. Since $\Delta$ is essentially the same as $E_{11}$, the left of Fig.~\ref{fig:optical}c shows the same trends as Fig.~\ref{fig:gap}c (here $E_{11}$ for $q=0$ is across the first transition in the optical range, not across the miniband). For $E_{22}$, however, the trend is reversed: $E_{22}$ increases for $q=2$ and decreases for $q=0$ and $q=1$. For the trends in the absolute transition energies, with different levels of tight-binding, see Refs.~\onlinecite{popov_PRB_04} and \onlinecite{popov_NJP_04}.

The analysis of $\varepsilon_2(\omega)$ excluded again armchair and small diameter tubes, armchair tubes for being insensitive for bending, and small diameter tubes for showing too unsystematic behavior. I further mention that properties with transverse-polarized light (dotted lines in Figs.\ref{fig:optical}a and \ref{fig:optical}b) are insensitive to bending; the tube cross-section remained, after all, rather spherical.

At this point it's pertinent to mention that these trends appear to contradict some earlier works. Reference~\onlinecite{liu_JPCM_06} reported bend-induced gap widening for $q=0$ and Refs.~\onlinecite{kane_PRL_97} and \onlinecite{yang_PRL_00} concluded that bending has no effect because---so the argument ran---opposite strains on opposite sides of the tube circumference cancel out. The very title of Ref.~\onlinecite{chibotaru_PRB_02} suggested bend-induced insulating gap for zigzag and chiral CNTs. While my results do agree on the insensitivity of armchair tubes towards bending,\cite{kane_PRL_97,yang_PRL_00} why are there contradictions in those other trends?


This question brings me to the third main result of this work: \emph{the correct trends are obtained only after full structural relaxation}. The absence of relaxation in previous works, therefore, explains the above contradictions. Indeed, if the calculations were repeated with a mere skilled estimate for the bent geometry, the trends regarding $q$-families turned out wrong. Even single-orbital tight-binding explains the trends correctly---if it's only given the properly optimized geometries.

What makes the relaxation so important, is that, \emph{due to anharmonic effects, the carbon-carbon bonds will, on average, stretch upon bending.} Pure stretching of CNTs, in turn, has been reported to yield the very same trends.\cite{yang_PRL_00} In the limit $\Theta \rightarrow 0$ bonds do behave harmonically and opposite strains on opposite sides of the tube do have a cancelling effect; note that all curves in Figs.\ref{fig:gap}c and \ref{fig:optical}c start with a zero slope, that is, anharmonicity emerges only in the second order in $\Theta$. As bending increases, the bonds in the outer part stretch progressively more than bonds in the inner part shrink. The simulated circumference-averaged strains increase as $\varepsilon_\text{avg}\sim 0.4\cdot \Theta^2$, yielding some $10^{-3}$ strains with $\Theta=5$~\%\ bending.



To conclude, the trends above ought to be correct and independent of any approximations of tight-binding, since the origin of the trends, the anharmonicity of carbon-carbon bonds, is so plausible and fundamental. Measuring these trends from individual tubes is feasible,\cite{wang_PRB_10} although care should be taken to avoid other distortions and to make the bending pure. More relevant, however, are still the general trends, not properties of individual tubes---because we still don't have full experimental control over nanotube chiralities.

I acknowledge A. Laakso for discussions, the Academy of Finland for funding, and the Finnish IT Center for Science (CSC) for computational resources.


\begin{thebibliography}{32}
\expandafter\ifx\csname natexlab\endcsname\relax\def\natexlab#1{#1}\fi
\expandafter\ifx\csname bibnamefont\endcsname\relax
  \def\bibnamefont#1{#1}\fi
\expandafter\ifx\csname bibfnamefont\endcsname\relax
  \def\bibfnamefont#1{#1}\fi
\expandafter\ifx\csname citenamefont\endcsname\relax
  \def\citenamefont#1{#1}\fi
\expandafter\ifx\csname url\endcsname\relax
  \def\url#1{\texttt{#1}}\fi
\expandafter\ifx\csname urlprefix\endcsname\relax\def\urlprefix{URL }\fi
\providecommand{\bibinfo}[2]{#2}
\providecommand{\eprint}[2][]{\url{#2}}

\bibitem[{\citenamefont{Ijima}(1991)}]{ijima_nature_91}
\bibinfo{author}{\bibfnamefont{S.}~\bibnamefont{Ijima}},
  \bibinfo{journal}{Nature} \textbf{\bibinfo{volume}{354}}, \bibinfo{pages}{56}
  (\bibinfo{year}{1991}).

\bibitem[{\citenamefont{Falvo et~al.}(1997)\citenamefont{Falvo, Clary, {Taylor
  II}, Chi, {Brooks Jr}, Washburn, and Superfine}}]{falvo_nature_97}
\bibinfo{author}{\bibfnamefont{M.~R.} \bibnamefont{Falvo}},
  \bibinfo{author}{\bibfnamefont{G.~J.} \bibnamefont{Clary}},
  \bibinfo{author}{\bibfnamefont{R.~M.} \bibnamefont{{Taylor II}}},
  \bibinfo{author}{\bibfnamefont{V.}~\bibnamefont{Chi}},
  \bibinfo{author}{\bibfnamefont{F.~P.} \bibnamefont{{Brooks Jr}}},
  \bibinfo{author}{\bibfnamefont{S.}~\bibnamefont{Washburn}}, \bibnamefont{and}
  \bibinfo{author}{\bibfnamefont{R.}~\bibnamefont{Superfine}},
  \bibinfo{journal}{Nature} \textbf{\bibinfo{volume}{389}},
  \bibinfo{pages}{582} (\bibinfo{year}{1997}).

\bibitem[{\citenamefont{Ko et~al.}(2004)\citenamefont{Ko, Pikus, Jiang, Jauss,
  Hollricher, and Tsukruk}}]{ko_APL_04}
\bibinfo{author}{\bibfnamefont{H.}~\bibnamefont{Ko}},
  \bibinfo{author}{\bibfnamefont{Y.}~\bibnamefont{Pikus}},
  \bibinfo{author}{\bibfnamefont{C.}~\bibnamefont{Jiang}},
  \bibinfo{author}{\bibfnamefont{A.}~\bibnamefont{Jauss}},
  \bibinfo{author}{\bibfnamefont{O.}~\bibnamefont{Hollricher}},
  \bibnamefont{and} \bibinfo{author}{\bibfnamefont{V.~V.}
  \bibnamefont{Tsukruk}}, \bibinfo{journal}{Appl. Phys. Lett.}
  \textbf{\bibinfo{volume}{85}}, \bibinfo{pages}{2598} (\bibinfo{year}{2004}).

\bibitem[{\citenamefont{Pan et~al.}(2002)\citenamefont{Pan, Zhang, and
  Nakayama}}]{pan_JAP_02}
\bibinfo{author}{\bibfnamefont{L.}~\bibnamefont{Pan}},
  \bibinfo{author}{\bibfnamefont{M.}~\bibnamefont{Zhang}}, \bibnamefont{and}
  \bibinfo{author}{\bibfnamefont{Y.}~\bibnamefont{Nakayama}},
  \bibinfo{journal}{J. Appl. Phys.} \textbf{\bibinfo{volume}{91}},
  \bibinfo{pages}{10058} (\bibinfo{year}{2002}).

\bibitem[{\citenamefont{Song et~al.}(2008)\citenamefont{Song, Ma, Ren, Zhou,
  Xie, Tan, and Sun}}]{song_APL_08}
\bibinfo{author}{\bibfnamefont{L.}~\bibnamefont{Song}},
  \bibinfo{author}{\bibfnamefont{W.}~\bibnamefont{Ma}},
  \bibinfo{author}{\bibfnamefont{Y.}~\bibnamefont{Ren}},
  \bibinfo{author}{\bibfnamefont{W.}~\bibnamefont{Zhou}},
  \bibinfo{author}{\bibfnamefont{S.}~\bibnamefont{Xie}},
  \bibinfo{author}{\bibfnamefont{P.}~\bibnamefont{Tan}}, \bibnamefont{and}
  \bibinfo{author}{\bibfnamefont{L.}~\bibnamefont{Sun}},
  \bibinfo{journal}{Appl. Phys. Lett.} \textbf{\bibinfo{volume}{92}},
  \bibinfo{pages}{121905} (\bibinfo{year}{2008}).

\bibitem[{\citenamefont{Wang et~al.}(2010)\citenamefont{Wang, Gupta, Huang,
  Vedala, Hao, Crespi, Choi, and Eklund}}]{wang_PRB_10}
\bibinfo{author}{\bibfnamefont{B.}~\bibnamefont{Wang}},
  \bibinfo{author}{\bibfnamefont{A.~K.} \bibnamefont{Gupta}},
  \bibinfo{author}{\bibfnamefont{J.}~\bibnamefont{Huang}},
  \bibinfo{author}{\bibfnamefont{H.}~\bibnamefont{Vedala}},
  \bibinfo{author}{\bibfnamefont{Q.}~\bibnamefont{Hao}},
  \bibinfo{author}{\bibfnamefont{V.~H.} \bibnamefont{Crespi}},
  \bibinfo{author}{\bibfnamefont{W.}~\bibnamefont{Choi}}, \bibnamefont{and}
  \bibinfo{author}{\bibfnamefont{P.~C.} \bibnamefont{Eklund}},
  \bibinfo{journal}{Phys. Rev. B} \textbf{\bibinfo{volume}{81}},
  \bibinfo{pages}{115422} (\bibinfo{year}{2010}).

\bibitem[{\citenamefont{Kutana and Giapis}(2006)}]{kutana_PRL_06}
\bibinfo{author}{\bibfnamefont{A.}~\bibnamefont{Kutana}} \bibnamefont{and}
  \bibinfo{author}{\bibfnamefont{K.~P.} \bibnamefont{Giapis}},
  \bibinfo{journal}{Phys. Rev. Let.} \textbf{\bibinfo{volume}{97}},
  \bibinfo{pages}{245501} (\bibinfo{year}{2006}).

\bibitem[{\citenamefont{Yakobson et~al.}(1996)\citenamefont{Yakobson, Brabec,
  and Bernholc}}]{yakobson_PRL_96}
\bibinfo{author}{\bibfnamefont{B.~I.} \bibnamefont{Yakobson}},
  \bibinfo{author}{\bibfnamefont{C.~J.} \bibnamefont{Brabec}},
  \bibnamefont{and} \bibinfo{author}{\bibfnamefont{J.}~\bibnamefont{Bernholc}},
  \bibinfo{journal}{Phys. Rev. lett.} \textbf{\bibinfo{volume}{76}},
  \bibinfo{pages}{2411} (\bibinfo{year}{1996}).

\bibitem[{\citenamefont{Pastewka et~al.}(2009)\citenamefont{Pastewka, Koskinen,
  Els\"asser, and Moseler}}]{pastewka_PRB_09}
\bibinfo{author}{\bibfnamefont{L.}~\bibnamefont{Pastewka}},
  \bibinfo{author}{\bibfnamefont{P.}~\bibnamefont{Koskinen}},
  \bibinfo{author}{\bibfnamefont{C.}~\bibnamefont{Els\"asser}},
  \bibnamefont{and} \bibinfo{author}{\bibfnamefont{M.}~\bibnamefont{Moseler}},
  \bibinfo{journal}{Phys. Rev. B} \textbf{\bibinfo{volume}{80}},
  \bibinfo{pages}{155428} (\bibinfo{year}{2009}).

\bibitem[{\citenamefont{Rochefort et~al.}(1999)\citenamefont{Rochefort,
  Avouris, Lesage, and Salahub}}]{rochefort_PRB_99}
\bibinfo{author}{\bibfnamefont{A.}~\bibnamefont{Rochefort}},
  \bibinfo{author}{\bibfnamefont{P.}~\bibnamefont{Avouris}},
  \bibinfo{author}{\bibfnamefont{F.}~\bibnamefont{Lesage}}, \bibnamefont{and}
  \bibinfo{author}{\bibfnamefont{D.~R.} \bibnamefont{Salahub}},
  \bibinfo{journal}{Phys. Rev. B} \textbf{\bibinfo{volume}{60}},
  \bibinfo{pages}{13824} (\bibinfo{year}{1999}).

\bibitem[{\citenamefont{Tang et~al.}(2010)\citenamefont{Tang, Guo, and
  Chen}}]{tang_JAP_10}
\bibinfo{author}{\bibfnamefont{C.}~\bibnamefont{Tang}},
  \bibinfo{author}{\bibfnamefont{W.}~\bibnamefont{Guo}}, \bibnamefont{and}
  \bibinfo{author}{\bibfnamefont{C.}~\bibnamefont{Chen}}, \bibinfo{journal}{J.
  Appl. Phys.} \textbf{\bibinfo{volume}{108}}, \bibinfo{pages}{026108}
  (\bibinfo{year}{2010}).

\bibitem[{\citenamefont{Nikiforov et~al.}(2010)\citenamefont{Nikiforov, Zhang,
  James, and Dumitric\v{a}}}]{nikiforov_APL_10}
\bibinfo{author}{\bibfnamefont{I.}~\bibnamefont{Nikiforov}},
  \bibinfo{author}{\bibfnamefont{D.-B.} \bibnamefont{Zhang}},
  \bibinfo{author}{\bibfnamefont{E.~D.} \bibnamefont{James}}, \bibnamefont{and}
  \bibinfo{author}{\bibfnamefont{T.}~\bibnamefont{Dumitric\v{a}}},
  \bibinfo{journal}{Appl. Phys. Lett.} \textbf{\bibinfo{volume}{96}},
  \bibinfo{pages}{123107} (\bibinfo{year}{2010}).

\bibitem[{\citenamefont{Dumitric{\u a} and James}(2007)}]{dumitrica_JMPS_07}
\bibinfo{author}{\bibfnamefont{T.}~\bibnamefont{Dumitric{\u a}}}
  \bibnamefont{and} \bibinfo{author}{\bibfnamefont{R.~D.} \bibnamefont{James}},
  \bibinfo{journal}{J. Mech. Phys. Solid} \textbf{\bibinfo{volume}{55}},
  \bibinfo{pages}{2206} (\bibinfo{year}{2007}).

\bibitem[{\citenamefont{Malola et~al.}(2008)\citenamefont{Malola, H\"akkinen,
  and Koskinen}}]{malola_PRB_08b}
\bibinfo{author}{\bibfnamefont{S.}~\bibnamefont{Malola}},
  \bibinfo{author}{\bibfnamefont{H.}~\bibnamefont{H\"akkinen}},
  \bibnamefont{and} \bibinfo{author}{\bibfnamefont{P.}~\bibnamefont{Koskinen}},
  \bibinfo{journal}{Phys. Rev. B} \textbf{\bibinfo{volume}{78}},
  \bibinfo{pages}{153409} (\bibinfo{year}{2008}).

\bibitem[{\citenamefont{Koskinen and Kit}(2010)}]{koskinen_PRL_10}
\bibinfo{author}{\bibfnamefont{P.}~\bibnamefont{Koskinen}} \bibnamefont{and}
  \bibinfo{author}{\bibfnamefont{O.~O.} \bibnamefont{Kit}},
  \bibinfo{journal}{Phys. Rev. Lett.} \textbf{\bibinfo{volume}{105}},
  \bibinfo{pages}{106401} (\bibinfo{year}{2010}).

\bibitem[{\citenamefont{Chang et~al.}(2007)\citenamefont{Chang, Okawa, Garcia,
  Majumdar, and Zettl}}]{chang_PRL_07}
\bibinfo{author}{\bibfnamefont{C.~W.} \bibnamefont{Chang}},
  \bibinfo{author}{\bibfnamefont{D.}~\bibnamefont{Okawa}},
  \bibinfo{author}{\bibfnamefont{H.}~\bibnamefont{Garcia}},
  \bibinfo{author}{\bibfnamefont{A.}~\bibnamefont{Majumdar}}, \bibnamefont{and}
  \bibinfo{author}{\bibfnamefont{A.}~\bibnamefont{Zettl}},
  \bibinfo{journal}{Phys. Rev. Lett.} \textbf{\bibinfo{volume}{99}},
  \bibinfo{pages}{045901} (\bibinfo{year}{2007}).

\bibitem[{\citenamefont{Porezag et~al.}(1995)\citenamefont{Porezag, Frauenheim,
  K\"ohler, Seifert, and Kaschner}}]{porezag_PRB_95}
\bibinfo{author}{\bibfnamefont{D.}~\bibnamefont{Porezag}},
  \bibinfo{author}{\bibfnamefont{T.}~\bibnamefont{Frauenheim}},
  \bibinfo{author}{\bibfnamefont{T.}~\bibnamefont{K\"ohler}},
  \bibinfo{author}{\bibfnamefont{G.}~\bibnamefont{Seifert}}, \bibnamefont{and}
  \bibinfo{author}{\bibfnamefont{R.}~\bibnamefont{Kaschner}},
  \bibinfo{journal}{Phys. Rev. B} \textbf{\bibinfo{volume}{51}},
  \bibinfo{pages}{12947} (\bibinfo{year}{1995}).

\bibitem[{\citenamefont{Koskinen and M\"akinen}(2009)}]{koskinen_CMS_09}
\bibinfo{author}{\bibfnamefont{P.}~\bibnamefont{Koskinen}} \bibnamefont{and}
  \bibinfo{author}{\bibfnamefont{V.}~\bibnamefont{M\"akinen}},
  \bibinfo{journal}{Computational Materials Science}
  \textbf{\bibinfo{volume}{47}}, \bibinfo{pages}{237} (\bibinfo{year}{2009}).

\bibitem[{hot()}]{hotbit_wiki}
\bibinfo{note}{Hotbit wiki \texttt{https://trac.cc.jyu.fi/projects/hotbit}}.

\bibitem[{\citenamefont{Elstner et~al.}(1998)\citenamefont{Elstner, Porezag,
  Jungnickel, Elsner, Haugk, Frauenheim, Suhai, and Seifert}}]{elstner_PRB_98}
\bibinfo{author}{\bibfnamefont{M.}~\bibnamefont{Elstner}},
  \bibinfo{author}{\bibfnamefont{D.}~\bibnamefont{Porezag}},
  \bibinfo{author}{\bibfnamefont{G.}~\bibnamefont{Jungnickel}},
  \bibinfo{author}{\bibfnamefont{J.}~\bibnamefont{Elsner}},
  \bibinfo{author}{\bibfnamefont{M.}~\bibnamefont{Haugk}},
  \bibinfo{author}{\bibfnamefont{T.}~\bibnamefont{Frauenheim}},
  \bibinfo{author}{\bibfnamefont{S.}~\bibnamefont{Suhai}}, \bibnamefont{and}
  \bibinfo{author}{\bibfnamefont{G.}~\bibnamefont{Seifert}},
  \bibinfo{journal}{Phys. Rev. B} \textbf{\bibinfo{volume}{58}},
  \bibinfo{pages}{7260} (\bibinfo{year}{1998}).

\bibitem[{\citenamefont{Popov and Henrard}(2004)}]{popov_PRB_04}
\bibinfo{author}{\bibfnamefont{V.~N.} \bibnamefont{Popov}} \bibnamefont{and}
  \bibinfo{author}{\bibfnamefont{L.}~\bibnamefont{Henrard}},
  \bibinfo{journal}{Phys. Rev. B} \textbf{\bibinfo{volume}{70}},
  \bibinfo{pages}{115407} (\bibinfo{year}{2004}).

\bibitem[{\citenamefont{Popov}(2004)}]{popov_NJP_04}
\bibinfo{author}{\bibfnamefont{V.~N.} \bibnamefont{Popov}},
  \bibinfo{journal}{New J. Phys.} \textbf{\bibinfo{volume}{6}},
  \bibinfo{pages}{17} (\bibinfo{year}{2004}).

\bibitem[{app()}]{approx2}
\bibinfo{note}{The angle $\alpha$ should also be an integer fraction of $2\pi$,
  and $\kappa$-points sampled discretely, not freely, between $[-\pi,\pi]$. The
  smallness of $\alpha$, however, makes these requirements insignificant.}

\bibitem[{\citenamefont{Bitzek et~al.}(2006)\citenamefont{Bitzek, Koskinen,
  G\"ahler, Moseler, and Gumbsch}}]{bitzek_PRL_06}
\bibinfo{author}{\bibfnamefont{E.}~\bibnamefont{Bitzek}},
  \bibinfo{author}{\bibfnamefont{P.}~\bibnamefont{Koskinen}},
  \bibinfo{author}{\bibfnamefont{F.}~\bibnamefont{G\"ahler}},
  \bibinfo{author}{\bibfnamefont{M.}~\bibnamefont{Moseler}}, \bibnamefont{and}
  \bibinfo{author}{\bibfnamefont{P.}~\bibnamefont{Gumbsch}},
  \bibinfo{journal}{Phys. Rev. Lett.} \textbf{\bibinfo{volume}{97}},
  \bibinfo{pages}{170201} (\bibinfo{year}{2006}).

\bibitem[{for()}]{force-note}
\bibinfo{note}{As a relevant technical detail, note that since $\alpha$ is
  small ($\sim 5\times 10^{-4}$~rad at minimum), the radial forces are
  effectively suppressed by this factor; for usual translation symmetry the
  force criterion corresponds to $0.02$~eV/\AA, or better. The strict force
  criterion is necessary.}

\bibitem[{\citenamefont{Huhtala et~al.}(2002)\citenamefont{Huhtala, Kuronen,
  and Kaski}}]{huhtala_CPC_02}
\bibinfo{author}{\bibfnamefont{M.}~\bibnamefont{Huhtala}},
  \bibinfo{author}{\bibfnamefont{A.}~\bibnamefont{Kuronen}}, \bibnamefont{and}
  \bibinfo{author}{\bibfnamefont{K.}~\bibnamefont{Kaski}},
  \bibinfo{journal}{Comput. Phys. Commun.} \textbf{\bibinfo{volume}{146}},
  \bibinfo{pages}{30} (\bibinfo{year}{2002}).

\bibitem[{\citenamefont{Arias and Arroyo}(2008)}]{arias_PRL_08}
\bibinfo{author}{\bibfnamefont{I.}~\bibnamefont{Arias}} \bibnamefont{and}
  \bibinfo{author}{\bibfnamefont{M.}~\bibnamefont{Arroyo}},
  \bibinfo{journal}{Phys. Rev. Lett.} \textbf{\bibinfo{volume}{100}},
  \bibinfo{pages}{085503} (\bibinfo{year}{2008}).

\bibitem[{\citenamefont{Yang and Han}(2000)}]{yang_PRL_00}
\bibinfo{author}{\bibfnamefont{L.}~\bibnamefont{Yang}} \bibnamefont{and}
  \bibinfo{author}{\bibfnamefont{J.}~\bibnamefont{Han}},
  \bibinfo{journal}{Phys. Rev. Lett.} \textbf{\bibinfo{volume}{85}},
  \bibinfo{pages}{154} (\bibinfo{year}{2000}).

\bibitem[{\citenamefont{Kane and Mele}(1997)}]{kane_PRL_97}
\bibinfo{author}{\bibfnamefont{C.~L.} \bibnamefont{Kane}} \bibnamefont{and}
  \bibinfo{author}{\bibfnamefont{E.~J.} \bibnamefont{Mele}},
  \bibinfo{journal}{Phys. Rev. Lett.} \textbf{\bibinfo{volume}{78}},
  \bibinfo{pages}{1932} (\bibinfo{year}{1997}).

\bibitem[{\citenamefont{Martin}(2004)}]{martin_book}
\bibinfo{author}{\bibfnamefont{R.~M.} \bibnamefont{Martin}},
  \emph{\bibinfo{title}{Electronic structure: Basic Theory and Practical
  Methods}} (\bibinfo{publisher}{Cambridge University Press},
  \bibinfo{year}{2004}).

\bibitem[{\citenamefont{Liu and Ding}(2006)}]{liu_JPCM_06}
\bibinfo{author}{\bibfnamefont{C.~P.} \bibnamefont{Liu}} \bibnamefont{and}
  \bibinfo{author}{\bibfnamefont{J.~W.} \bibnamefont{Ding}},
  \bibinfo{journal}{J. Phys.:Condens. Matter} \textbf{\bibinfo{volume}{18}},
  \bibinfo{pages}{4077} (\bibinfo{year}{2006}).

\bibitem[{\citenamefont{Chibotaru et~al.}(2002)\citenamefont{Chibotaru, Bovin,
  and Ceulemans}}]{chibotaru_PRB_02}
\bibinfo{author}{\bibfnamefont{L.~F.} \bibnamefont{Chibotaru}},
  \bibinfo{author}{\bibfnamefont{S.~A.} \bibnamefont{Bovin}}, \bibnamefont{and}
  \bibinfo{author}{\bibfnamefont{A.}~\bibnamefont{Ceulemans}},
  \bibinfo{journal}{Phys. Rev. B} \textbf{\bibinfo{volume}{66}},
  \bibinfo{pages}{161401} (\bibinfo{year}{2002}).

\end{thebibliography}

\end{document}